\documentclass[useAMS,usenatbib]{mn2e}

\usepackage{graphicx}
\usepackage{caption}
\usepackage{cleveref}
\title[On the morphology of bar dust lanes]{On the morphology of dust lanes in galactic bars} 
\author[L. S\'anchez-Menguiano et al.]{L. ~S\'anchez-Menguiano,$^{1,2}$ I. ~P\'erez,$^2$ A. ~Zurita,$^2$ I. ~Mart\'inez-Valpuesta,$^{3,4}$
\newauthor
J. A. L. ~Aguerri,$^3$ S. F. ~S\'anchez,$^5$ S. ~Comer\'on$^{6,7}$ and S. ~D\'iaz-Garc\'ia$^{6}$\\
$^1$Instituto de Astrof\'isica de Andaluc\'ia (CSIC), Glorieta de la Astronom\'ia s/n, Aptdo. 3004, E-18080 Granada, Spain\\
$^2$Dpto. de F\'isica Te\'orica y del Cosmos, Universidad de Granada, Facultad de Ciencias (Edificio Mecenas), E-18071 Granada, Spain\\
$^3$Instituto de Astrof\'isica de Canarias, E-38205 La Laguna, Tenerife, Spain\\
$^4$Universidad de La Laguna, Dpto. Astrof\'isica, E-38206 La Laguna, Tenerife, Spain\\
$^5$Instituto de Astronom\'ia, Universidad Nacional Aut\'onoma de M\'exico, A.P. 70-264, 04510, M\'exico, D.F.\\
$^6$University of Oulu, Astronomy Division, Department of Physics, P.O. Boc 3000, FI-90014, Finland\\
$^7$Finnish Centre of Astronomy with ESO (FINCA), University of Turku, V\"ais\"al\"anti 20, FI-21500, Piikki\"o, Finland}

\begin{document}

\date{Accepted 2015 April 7. Received 2015 April 6; in original form 2014 December 9}
\maketitle
\begin{abstract}
The aim of our study is to use dynamical simulations to explore the influence of two important dynamical bar parameters, bar strength and bar pattern speed, on the shape of the bar dust lanes. To quantify the shape of the dust lanes we have developed a new systematic method to measure the dust lane curvature. Previous numerical simulations have compared the curvature of bar dust lanes with the bar strength, predicting a relation between both parameters which has been supported by observational studies but with a large spread. We take into account the bar pattern speed to explore, simultaneously, the effect of both parameters on the dust lane shape. To that end, we separate our galactic bars in fast bars $\left(1 < \mathcal{R} < 1.4 \right)$ and slow bars $\left(\mathcal{R} > 1.4 \right)$, obtaining, as previous simulations, an inverse relation between the dust lane curvature and the bar strength for fast bars. For the first time, we extend the study to slow bars, finding a constant curvature as a function of the bar strength. As a result, we conclude that weak bars with straight dust lanes are candidates for slow bars. Finally, we have analysed a pilot sample of ten S$^4$G galaxies, obtaining dust lane curvatures lying within the range covered by the simulations.
\end{abstract}


\section{Introduction}
Bars are common morphological features among spiral galaxies. Roughly 30-40\% of the spiral galaxies have a pronounced bar in optical wavelengths and if we take into account weaker bars, the fraction rises up to 60\% (\citealp{vacu, sell, marinova}; \citealp*{barazza}; \citealp{sheth}; \citealp*{aguerri2009}; \citealp{nair, masters}). Kinematic data shows the presence of strong non-circular gas motions in bars \citep[e.g.~][]{huntley, zurita}, which indicates that the bar constitutes a major non-axisymmetric component of the galaxy mass distribution \citep{sell}. 

Stellar bars are thought to be a key mechanism in the dynamical evolution of disc galaxies. For example, they are able to contribute to the redistribution of matter in the galaxy by exchanging angular momentum with the disc, bulge and halo (e.g. \citealt{deb1998, deb, lia2003,  cheung2013} and reviews by \citealt{kormendy}; \citealt{lia2013} and \citealt{sell2014}).

Galactic bars are characterised dynamically by three main parameters: length, strength\footnote{Parameter that quantifies the effectiveness to which a bar potential influences the motions of stars.} and pattern speed. Several methods have been proposed to measure them. Determining the bar length is not entirely trivial. For instance, early-type galaxies do not usually have obvious spiral structure that demarcates the bar end. In other cases, the presence of a large bulge may also complicate the measurement. For example, \citet*{aguerri} applied four different criteria to determine the bar length, which leads to variations of $\sim$15-20\% in measured length\footnote{Based on the mean values and standard deviations from their Table 4.}. Visual inspection of galaxy images \citep[e.g.~][]{martin}, ellipse fitting of the isophotes by locating the maximum in the ellipticity profile \citep[e.g.~][]{wozniak1991} or by looking for variations of the position angle \citep[e.g.~][]{wozniak1995,aguerri2009} and Fourier analysis of the surface brightness \citep[e.g.~ \citealp*{ohta,aguerri1998};][]{aguerri2000a} are among the several methods proposed for measuring the bar length. For details on different techniques and definitions used to determine the bar length see, for instance, \citet{michel2006} and \citet{gadotti2007}. Studies have obtained a typical value for the bar length of 6-8 kpc, being bars longer by a factor of $\sim$ 2.5 in early-type disc galaxies than in late-type disc galaxies \citep{erwin}.

Regarding bar strength, different parameters have been proposed to describe it.  One of the simplest is the deprojected bar ellipticity \citep{martin}. \citet{combes} proposed a parameter $Q_{\rm b}$ which represents the maximum bar torque applied to a gaseous material in orbital motion relative to its specific kinetic energy. This is one of the most frequently used parameters to describe bar strength. Fourier techniques are also commonly used to measure the bar torque (e.g.~ \citealp{aguerri2000a}; \citealp*{buta2,inma2006}).

The bar pattern speed is one of the most defining dynamical parameters. It can be parametrised by a distance-independent parameter $\mathcal{R} = R_{\mathrm{CR}}/R_{\mathrm{bar}}$, where $R_{\rm{CR}}$ is the Lagrangian/corotation radius and $R_{\mathrm{bar}}$ is the bar semimajor axis. Bars that end near corotation $\left(1 < \mathcal{R} < 1.4 \right)$ are considered as fast bars and shorter bars $\left(\mathcal{R} > 1.4 \right)$ as slow bars \citep{deb}. For  $\mathcal{R} < 1$, orbits are elongated perpendicular to the bar and self-consistent bars cannot exist \citep{contopoulos}. The bar pattern speed is also difficult to measure. Several methods are usually used to determine it, but the uncertainties  in the analysis give inconclusive results \citep{knapen99}. The most direct determination is by applying the Tremaine-Weinberg method, based on the continuity equation \citep{TW}, but it requires imaging data with high signal-to-noise and, therefore, long integration times, which limits its application to a restricted number of candidates \citep[e.g.~ \citealp*{deb2002,gerssen};][]{aguerri,treut}. \citet{aguerri2015} has applied this method to a larger sample of galaxies taking advantage of integral field spectroscopy data from the CALIFA survey \citep{califa}, not finding differences of pattern speed with morphological types. Recent work \citep{font2014} has presented a new direct method to locate resonances in a disc using 2-D gas information. It requires high velocity and spatial resolution of ionised gas data. They find that most of the analysed bars are consistent with being fast and present a hint of pattern speed segregation with morphological types.

These properties of bars determine their influence in the galactic dynamics and produce or modify some morphological features. Due to the difficulties in determining these bar parameters, it is highly desirable to find indirect methods to derive them, like the use of morphological information of the bar. Previous studies have derived the bar pattern speed matching observational features with resonances, for instance using rings as indicators \citep[e.g.~][]{buta86}, even for galaxies at high redshift \citep*{perez2012}. One of the most remarkable features of bars is the presence of dust lanes along them that extend from the nuclear region into the spiral arms. Bar dust lanes have been studied from numerical simulations (e.g.~\citealp{albada,lia,patsis2000}; \citealp*{patsis2010,kim2012}), as well as observationally \citep[e.g.~ \citealp*{knapen};][]{marshall2008,comeron}. These show different morphologies; from completely straight, although they can sometimes curl around the centre, to curved, with the concave sides towards the major axis. 

Studies of gas flows in bars (mainly theoretically from N-body simulations, e.g. \citealp{fux}; \citealp*{perez2004}; \citealp{perez2008}) have shown that dust lanes are sites of high gas and dust density, and therefore, of high absorption of the starlight. Numerical simulations \citep{lia,patsis2010} have revealed the existence of shocks at the positions of these narrow lanes. The existence of shocks along the dust lanes has been observationally shown from the analysis of gas kinematics of strongly barred galaxies \citep[e.g.~][]{downes, mundell1999, zurita}. These shocks are tracing the channels by which the gas flows towards the galaxy centres where it can contribute to the central mass concentration, for example producing star formation, pseudo-bulges formation or AGN activity \citep[e.g.~ \citealp*{shlosman,oh2012};][]{wang2012,lee2012}.

Numerical simulations have pointed out a relation between the shape of the dust lanes and the bar strength, where stronger bars show straighter dust lanes, and between the position of the dust lanes and the bar pattern speed, where dust lanes that are offset from the bar major axis appear for a limited range of bar pattern speeds, below which the dust lanes are `centred' \citep{lia}. There have been a few attempts to observationally find this relation between dust lane curvature and bar strength \citep{knapen, comeron}, but without taking into account the bar pattern speed. Even though \citet{knapen} confirmed a correlation between bar strength and dust lane curvature, \citet{comeron}, using a larger sample and improving the number statistics from \citet{knapen}, showed a large spread in the data and concluded that, though bar strength set an upper limit to the dust lane curvature allowed, this parameter by itself did not determine the dust lane curvature.

In this paper we use dynamical simulations to explore the effect of bar strength and bar pattern speed on the shape of the bar dust lanes. To carry out this study we have developed a new methodology to characterise the dust lane shapes based on the mathematical definition of curvature on a set of simulated galaxies. The final aim of the work would be to derive information about the underlying dynamics of bars by a simple analysis of morphological features, with easier measurements.

The structure of the paper is organised as follows. In Section 2, we provide a description of the simulations used in this study. In Section 3, we explain the methodology developed to measure the bar dust lane curvature. In Section 4, we present the results found in this work and analyse the dependency of the curvature on bar parameters. Finally, in Section 5 we discuss the results and compare them with previous studies in this topic. 

\section{Dynamical simulations}
To study the dynamical response of the gas in the presence of a galactic bar we used an updated version of the FTM 4.4 code from \citet{heller}, a three-dimensional hybrid SPH/N-body code. We ran a set of 238 simulations, which were performed with $10^5$ isothermal, non self-gravitating, collisional particles. These are the same simulations used in \citet{comeron}. 

\begin{figure*}
\centering
\includegraphics[width=\textwidth]{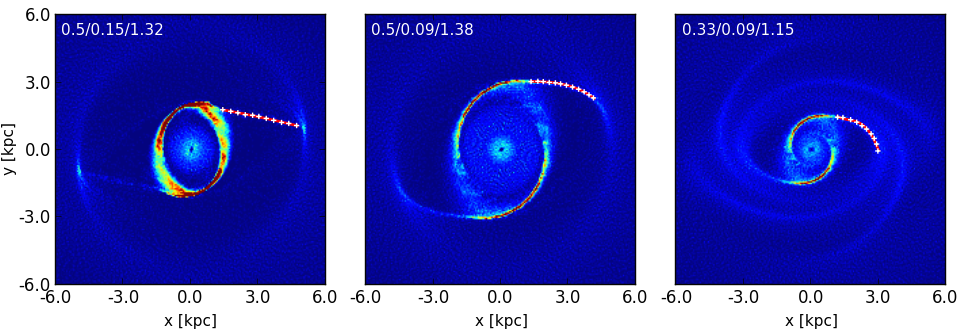}
\caption{Gas density maps of the bar region for three simulated galaxies with one of the dust lanes marked (white crosses). Bar major axis is parallel to the horizontal in all frames. We have also plotted the dust lane fitting using a third degree polynomial (red line), see text for details. Ellipticity, bar strength and the ratio of corotation radius to bar semimajor axis ($\mathcal{R}$) are given in this order in the upper left-hand corner of each frame. From left to right, the dust lane curvatures (as explained in Sect. \ref{sec:curv}) are 1.0, 1.3 and 2.3.}
\label{fig:tracing}
\end{figure*}

The potential used in the simulations consists of three axisymmetric components associated to the galaxy disc, bulge and halo, each one modelled by a Miyamoto-Nagai profile, and a non-axisymmetric component identified with the bar, modelled by a Ferrers potential \citep{ferrers} with $n=1$, being $n$ the degree of  the central density concentration. The halo and disc masses are constant for all the simulations, but we vary the bulge to total mass ratio, $B/(B+D)$, by increasing the bulge mass. We have used in the simulations two different values of the bar mass, being one bar a 50\% more massive than the other. For more details about the parameter values used in the potential models, see \citet{comeron}.

Regarding the bar pattern speed, we use $\Omega_b$ equal to 10, 20, 30 and 40 km s$^{-1}$ kpc$^{-1}$, placing the corotation radius outside the bar and obtaining values for $\mathcal{R}$ between 1.0 and 3.34. Bars with $R_{\mathrm{CR}}/R_{\mathrm{bar}} > 3.0$ are very rare in nature. Only one is found in the compilation made by \citet{rautiainen}, with $\mathcal{R} = 3.43 \pm 0.95$, and even in that case the error may still place the value below the limit of 3.0. For this reason, we have discarded those simulations in which corotation radius is more than three times the bar radius. After this, the number of simulations drops to 175.

We derive analytically the bar strength of the simulated galaxies by using the parameter proposed by \citet{combes}, $Q_{\rm b}$, which is given by the maximum over $r$ of

\begin{equation}
Q_{\rm b} =\frac{F_{\rm T}^{\rm max}(r)}{< F_{\rm R} (r)>} = \frac{\frac{1}{r} \left| \frac{\partial \Phi (r, \phi)}{\partial \phi}\right|_{\rm max}}{<\frac{\rm{d} \Phi_0 (r)}{\rm{d} r}>},
\end{equation}
where $r$ is the galactocentric distance, $F_{\rm T}^{\rm max}$ is the azimuthal maximum of the absolute tangential force in the bar region, $< F_{\rm R} (r)>$ the mean axisymmetric radial force, $\Phi (r, \phi)$ the in-plane gravitational potential and $\Phi_0 (r)$ its axisymmetric component.

\section{Measurement of the bar dust lane curvature}
\label{sec:curv}

A way to characterise and quantify the shape of bar dust lanes is by measuring its curvature. As dust lanes are related to gas shocks, we measure their curvature in our set of simulated galaxies by deriving the curvature of the gas density enhancements, using a systematic and easily reproducible method, explained below. The range we cover to measure the dust lane curvature starts outside the bar inner region, defined by the nuclear ring associated to the ILR in some cases, or a ring without closure in others, and finishes at the beginning of spiral arms, corresponding to a kink in the dust lane direction. This measurement was performed after two bar rotations in the simulations. Well-defined measurable dust lanes appeared in 126 simulations. This final sample does not exactly match the sample in \citet{comeron}, which comprises 88 simulations. Regrettably, it is not possible to access the sample presented in \citet{comeron}, and therefore, we cannot check the source of the discrepancy in the number of galaxies.

To measure the curvature we visually trace one of the dust lanes in each simulation (in our simulations both dust lanes are symmetrical) making use of SAOImage ds9\footnote{An astronomical imaging and data visualisation application developed by Smithsonian Astrophysical Observatory \citep{ds9}.}. Figure \ref{fig:tracing} shows the outline of a dust lane in three simulated galaxies. Then we fit the points tracing the bar dust lane with a third degree polynomial (Figure \ref{fig:tracing}).
To check the goodness of the fit we calculate the medium residuals of the fit for each galaxy, which is a measure of the discrepancy between the data and the model. Performing this test, the mean value of the residuals for all galaxies is of 0.2 pixels\footnote{Taking the unit length of simulations to 10 kpc, each pixel corresponds to 60 pc.}, with a standard deviation of 0.1 pixels, which reflects that a third degree polynomial fits quite well the data. In the particular case of the simulated galaxies shown in Figure \ref{fig:tracing}, from left to right the maximum residual for each fitting is 0.43, 0.16 and 0.69 pixels, with a mean residual along the fitting of 0.18, 0.08 and 0.38 pixels, respectively.

From the fitting of the bar dust lanes we now measure their curvature. The curvature of a plane curve gives information on how fast its tangent vector changes direction as you travel along the curve, and this is exactly what we want to know about the bar dust lane shape. If the dust lane keeps close to the same direction (that is, the dust lane is straight), the unit tangent vector changes very little and the curvature is small (the opposite if the dust lane undergoes a tight turn). To measure bar dust lane curvature we use the mathematical definition of curvature for a plane curve written in cartesian coordinates in the form $y = f(x)$

\begin{equation}
\kappa \equiv \frac{\mathrm{d} \phi}{\mathrm{d}s} = \frac{| y''(x)|}{\left(1+\left(y'(x)\right)^2\right)^{3/2}}
\end{equation}
where $\phi$ is the tangential angle and $s$ is the arc length. It has units of inverse distance. For more details about the mathematical deduction of the curvature of a function see, for example, \citet{curvature}. 

We should not forget that this mathematical curvature is not a constant function. Therefore, we have adopted as `curvature of a bar dust lane' the absolute mean value of the curvature along the bar dust lane. This parameter is zero for straight bar dust lanes and progressively increases as these deviate from straight lines.

Besides, we should also notice that the curvature, as defined before, is a measurement that depends on the absolute bar size. As in the simulations this parameter is constant, we do not have to deal with this issue. However, to apply this method to observational data, we need to `normalise' the curvature to obtain a dimensionless parameter which takes into account the size of the bars, multiplying the mean curvature by the bar semimajor axis. 

To show that the method is viable and easy to implement on observational studies, we have applied it to a pilot sample. This sample comprises the galaxies from \citet{comeron} for which there are measurements of their bar pattern speed in \citet{rautiainen}. $Q_{\rm b}$ values are taken from D\'iaz-Garc\'ia et al. (in prep), who inferred the gravitational potential from the 3.6 $\mu$m imaging of the Spitzer Survey of Stellar Structure in Galaxies sample (S$^4$G)\footnote{http://www.cv.nrao.edu/$\sim$ksheth/S4G/} following the recipe from \citet{lauri} and \citet{salo2010}. To obtain the curvature values, we have followed the same procedure as for the simulations, using $g$ or colour-index $g-r$ images, proxies for dust distribution, from SDSS when available and otherwise, other optical images from the NASA / IPAC Extragalactic Database (NED). Table \ref{tab:objects} shows the names of the galaxies included in the study, as well as some of their properties and the derived normalised curvatures, for both bar dust lanes when possible, and just for one when the other is not well-defined.

\begin{table}
\tabcolsep=0.19cm
   \caption{Properties of the real galaxies used in the study.}
   \begin{tabular}{cccccccc} 
      \hline
      Name  & i$_d$ & PA$_d$ &  $\mathcal{R}$ & a$_{b}$ & $\varepsilon_{b}$ & Q$_{b}$ & $\kappa$\\
       \scalebox{.8}{(NGC)} & \scalebox{.8}{(deg)} & \scalebox{.8}{(deg)} &  & \scalebox{.8}{(kpc)} &  &  & \\
      \hline
      3504 & 20.93 & 138.3 & 1.19 &  4.20 & 0.63 & 0.26 & 1.35/1.54\\ 
      4123 & 39.45 & 119.9 & 1.17 & 4.52 & 0.65 & 0.57 & 0.77/0.77\\
      4303 & 24.77 & 153.0 & 1.70 & 6.36 & 0.68 & 0.42 & 1.40/1.14\\
      4314 & 22.78 & 42.7 & 0.99 & 6.63 & 0.67 & 0.45 & 0.20\\
      4457 & 17.64 & 69.7 & 0.98 & 2.60 & 0.35 & 0.09 & 2.37\\
      4548 & 31.46 & 144.2 & 1.26 & 5.88 & 0.61 & 0.28 & 0.46/1.29\\
      4579 & 43.20 & 96.5 & 1.46 & 6.05 & 0.49 & 0.18 & 1.70\\
      5921 & 39.48 & 120.3 & 1.25 & 5.82 & 0.80 & 0.36 & 0.12/1.22\\
      7552 & 15.63 & 26.0 & 0.99 & 4.84 & 0.66 & 0.36 & 1.16/0.34\\
      7723 & 44.30 & 47.1 & 1.31 & 2.81 & 0.50 & 0.27 & 0.77\\

      \hline
   \end{tabular}
   \label{tab:objects}
   
      \smallskip
   The table shows the galaxy name (Column 1), the galaxy disc inclination and disc PA (Columns 2 and 3) from \citet{s4g}, the bar pattern speed parametrised by $\mathcal{R}$ from \citet{rautiainen} (Column 4), the deprojected bar semimajor axis and ellipticity (Columns 5 and 6) from \citet{s4g}, the bar strength (Column 7) from D\'iaz-Garc\'ia et al. (in preparation) and the dust lane curvature measured as defined in Section \ref{sec:curv} (Column 8).
\end{table}

\section{Results}

We want to assess the influence of bar strength and bar pattern speed on the curvature of bar dust lanes. For that purpose, we have developed in this work a new methodology to quantitatively measure bar dust lane curvatures (see Section \ref{sec:curv}). 

\begin{figure}
\centering
\includegraphics[width=0.48\textwidth]{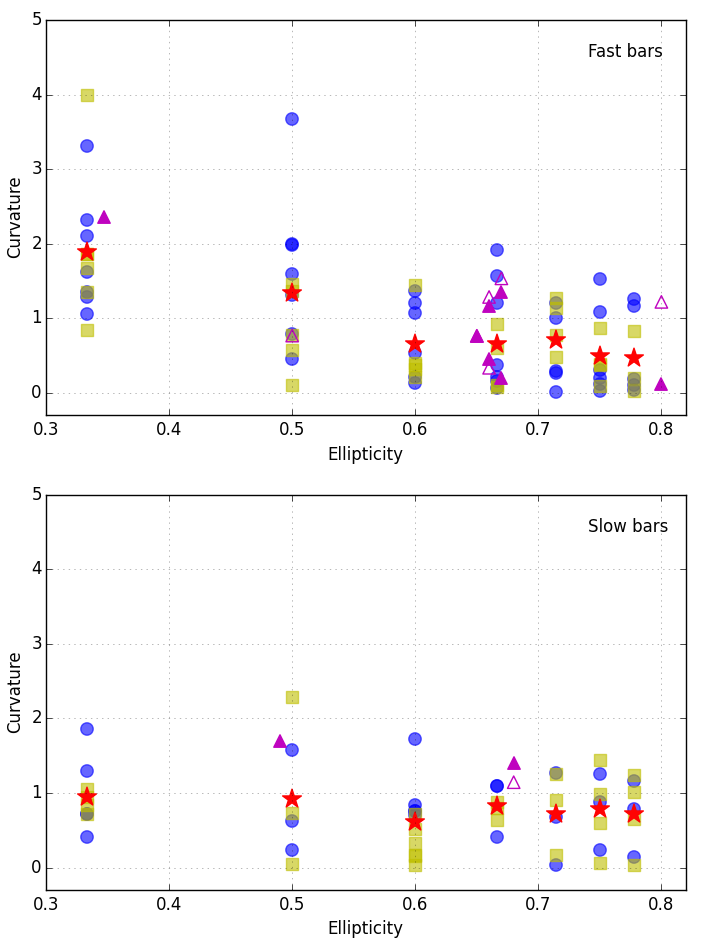}
\caption{Dust lane curvatures as a function of the bar ellipticity $\epsilon$ for fast bars, i.e., $1 < \mathcal{R} < 1.4 $ (upper) and for slow bars, i.e., $\mathcal{R} > 1.4 $ (bottom). The red stars mark the medium value of curvature for each value of the ellipticity. The blue circular markers are the values for galaxies with a bar mass of $M_b = 1.0 \times 10^{10} \mathrm{M}_{\odot}$ and the yellow squares for $M_b = 1.5 \times 10^{10} \mathrm{M}_{\odot}$. The purple triangles correspond to the values for the real galaxies.}
\label{fig:results_el}
\end{figure}

In Figure \ref{fig:results_el} we show the derived dust lane curvatures as a function of the bar ellipticity distinguishing between fast ($1 < \mathcal{R} < 1.4$) and slow ($\mathcal{R} > 1.4$) bars. The red stars are the mean curvature values for each ellipticity. The blue circular markers are the values for galaxies with a bar mass of $M_b = 1.0 \times 10^{10} \,\mathrm{M}_{\odot}$ and the yellow squares for $M_b = 1.5 \times 10^{10} \,\mathrm{M}_{\odot}$. Finally, the purple triangles are the values for the real galaxies included in the study. In the cases in which it has been possible to measure both dust lanes, they are distinguished using both filled and unfilled markers.

Regarding the mean curvature values, it is clear that for fast bars the curvature is smaller for higher ellipticity values, and slow bars show similar values of the mean curvature for all ellipticity values. There seems to be a lack of points for fast bars at low ellipticities and low curvatures, this effect is not so evident for slow bars. The most striking result from this figure is the absence of high curvature points for slow bars. In fact, the average curvature value for slow bars at the lowest ellipticity is 0.95 while the value for fast bars is 1.90. We show later in this section that the distribution of points for both fast and slow bars is clearly different. As it can be seen from Figure \ref{fig:results_el}, the values for the real galaxies are within the range of parameters covered by the simulations.

\begin{figure}
\centering
\includegraphics[width=0.48\textwidth]{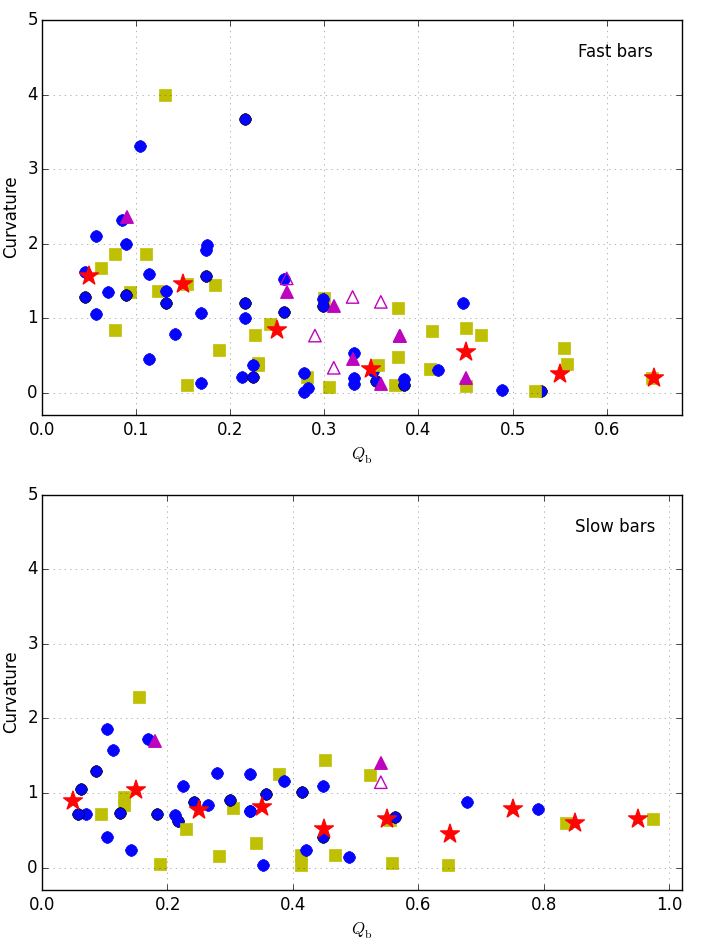}
\caption{Dust lane curvatures as a function of the bar strength $Q_b$ for fast bars, i.e., $1 < \mathcal{R} < 1.4 $ (upper) and for slow bars, i.e., $\mathcal{R} > 1.4 $ (bottom). The red stars are the medium values of the curvature for ranges of the bar strength of 0.1. The blue circular markers are the values for galaxies with a bar mass of $M_b = 1.0 \times 10^{10} \mathrm{M}_{\odot}$ and the yellow squares for $M_b = 1.5 \times 10^{10} \mathrm{M}_{\odot}$. The purple triangles correspond to the values for the real galaxies.}
\label{fig:results_qb}
\end{figure}

Figure \ref{fig:results_qb} presents the curvature against the bar strength, again distinguishing between fast (top panel) and slow bars (bottom panel). The marker symbols and colours used are the same as in the previous figure, being the red stars the mean curvature values for bar strength bins of 0.1. 

We observe similar trends to those found in Figure \ref{fig:results_el}, indicating that, in this case, the ellipticity is a good proxy for bar strength. Previous studies comparing bar strength and ellipticity for a large sample of galaxies have reached a similar conclusion \citep*[e.g.~][]{lauri2}. For slow bars, we have almost constant curvature value for the whole range of bar strengths while for fast bars we found higher curvature values at low bar strength. As shown in Figure \ref{fig:results_el}, there is a lack of low curvature points for fast bars at low bar strengths and a deficit of high curvature points for slow bars. Also in this case, the values for the real galaxies are within the range of parameters covered by the simulations.

We perform a two-sample Kolmogorov-Smirnov test to check if the differences found in the results between fast and slow bars are statistically significant, obtaining a P-value of 12\%. The significance level of the K-S test is 5\%, meaning that values below this limit come from different distributions. Therefore, the two complete samples seem to have similar distributions. However, if we restrict the comparison range to weak bars, with $Q_{\rm b} \leq 0.3$, where we observe differences between both distributions, the P-value is now 2\%. This result indicates that slow and fast bars have a different distribution of curvature for this range of $Q_{\rm b}$ values. At low bar strengths, the statistics of our sample is  considerably reduced ($\approx$ 30 elements). Because of this we have also performed an Anderson-Darling test, more appropriate for small samples, obtaining a P-value of 5\%, reinforcing the observed trends. 

\section{Discussion and conclusions}

In this work we have used numerical simulations to explore the effect of two bar parameters, bar strength and bar pattern speed, on the bar dust lane shape. To characterise the shape of dust lanes, we have developed a new method based on the mathematical curvature of a plane curve, providing a measurement which quantifies this shape, and we have then used these measurements to study the influence of the bar parameters. To do that, we have analysed 126 barred simulations, covering bar strengths from 0 to 1 and $\mathcal{R}$ between 1.0 and 3.0 for two bar masses. We have also applied this method to a set of observational data.

\citet{lia} studied the shape of bar dust lanes and its dependency on a set of model parameters, like the Lagrangian radius, the axial ratio of the bar and the bar quadrupole momentum $\mathcal{Q}_m$ as a measurement of the bar mass. However, the characterisation she made of the shape of the dust lanes was purely visual, not providing a method to quantify its curvature and it is therefore difficult to directly apply on observations. 

\citet{lia}, analysing simulations for fast bars, deduced that curved dust lanes result from weak bars and straight ones originate from strong bars. Our simulation results confirm this prediction. We extend the study to slow bars and find that the curvature does not vary with bar strength. \citet{lia} also included in the study the influence of the bar pattern speed, but more related to the position of the dust lanes than to the shape, which is the purpose of this work. She found a tight range of Lagrangian radii, which corresponds to fast bars in the distinction we have made, that leads to dust lanes that are `offset' from the bar major axis. For smaller values, corresponding to slow bars, the dust lanes are `centred' and for larger values they turn their convex sides towards the bar major axis, corresponding to unrealistic dust lanes.

A methodology similar to the one presented here, is used by \citet{knapen} and \citet{comeron} on real galaxies. These works also developed a way to quantify the curvature of dust lanes using real galaxies. Their method is based on measurements of the change in the angle of the tangent to the curved dust lanes in the range where the curvature is constant, \citet{knapen} expressing the result in units of degrees per kpc and \citet{comeron} multiplying that angle by the angular radius at which the torque is maximal, to take into account the size of the host bars. This method has the advantage of being simple and rather straightforward, but it also presents a number of drawbacks. First, it depends a lot on which part of the dust lane is selected to measure the curvature; and therefore, it is not easy to replicate their results. And second, they also assume a constant curvature for the bar dust lanes, which is not always the case (52\% of our simulations show dust lanes with non-constant curvature).

Our methodology does not assume any a priori shape for the dust lanes and the results are easy to reproduce. We select the beginning and end of the dust lanes in such way that we avoid morphological structures, such as nuclear rings and spiral arms, that may be masking the actual dust lane curvature. As we discussed in Sect. \ref{sec:curv}, we define the end of the dust lane as a kink in the dust lane direction which corresponds to the beginning of the spiral arms. For most of the galaxies, this parameter ranges within a similar galactocentric radius, indicating that our method measures dust lanes close to the bar ends in almost all cases. We have checked the effect of shortening the dust lanes in the curvature determination to test those cases where the dust lanes do not reach the end of the bar. We have done this test on 40 galaxies obtaining that the mean curvature is only changed by 19\% on average.

\citet{knapen} results agree with the correlation between the dust lane curvature and the bar strength found by \citet{lia}. However, \citet{comeron}, using a larger sample, obtained a large spread in the relation that made them state that bar strength by itself did not determine the dust lane curvature. 

None of these observational studies take into account the bar pattern speed as an important parameter influencing the bar dust lane shapes. However, numerical simulations \citep[e.g.~][]{albada,lia,fux,perez2004,perez2008} have shown that the location of gas shocks is linked to the bar potential and the pattern speed. We take a step forward adding to the distribution of bar strengths a segregation with bar pattern speed, dividing the sample in fast and slow bars, to find a relation between both parameters and dust lane curvature. The result obtained in the performed Kolmogorov-Smirnov test supports our assumption that the bar pattern speed has influence in the shape of bar dust lanes. We find that this inverse relation between the dust lane curvature and the bar strength only holds for fast bars, obtaining for slow bars approximately a constant curvature for all the values of the bar strength. 

The distribution of points shown in Figure \ref{fig:results_qb} indicates a larger probability of finding slow bars among galaxies with weak bars and straight dust lanes. This is an interesting effect to be checked observationally.

We have tested this method on a set of S$^4$G galaxies for which we have pattern speed and bar strength measurements. We have followed the same procedure as described for the simulations. Despite the fact that the sample is not large enough to confirm the trends found here, the obtained values lie within the range of values covered by the simulations. 

Although all parameters have been measured in a similar way for both simulations and observations, the halo potential is not included in the measurement of the bar strength in the S$^4$G galaxies. In principle, this could modify the results since the dust lane morphology depends on the whole galaxy potential. However, the analysed sample is comprised of high surface brightness galaxies for which is commonly assumed a maximum disc, implying that the contribution of the dark halo in the optical disc is minimum (e.g. ~\citealp{sancisi, broeils}; \citealp*{weiner}; \citealp{perez2004}). Because of this we do not expect the results to be substantially changed. Furthermore, the fact that the bar strength follows the ellipticity trend also reinforces this assumption. 

In summary, we have developed a method to measure dust lane curvatures that is robust and easily applicable to observational data. A future work using a larger sample of galaxies with measurements of bar strength and pattern speed is necessary to test the trends found with the simulations.

\section*{Acknowledgements}

We acknowledge financial support from the Spanish {\em  Ministerio de  Econom\' ia y Competitividad} via grants AYA2011-24728 and AYA2012-31935, and from  the `Junta de Andaluc\' ia' local government through the FQM-108 project. We also acknowledge support to the DAGAL Network from the People Programme (Marie Curie Actions) of the European Union's Seventh Framework Programme FP7/2007- 2013/ under REA grant agreement number PITN-GA-2011-289313.

\bibliography{bibliography}

\end{document}